\documentclass[aps,pra,twocolumn,groupedaddress]{revtex4-1}
 \usepackage{graphicx}

\begin{document}


\title{Relativistic tunneling through two "transparent" successive barriers}


\author{Massimo Germano}

\affiliation{University of Rome "La Sapienza", S.B.A.I. Department (Basic and Applied Sciences for Engineering), via Antonio Scarpa 16, 00161 Rome, Italy}


\date{\today}

\begin{abstract}
In the case of tunneling of relativistic particles, differently from the nonrelativistic case, a limit of "transparent" barrier can also lead to an apparent "superluminal" behavior when considering the phase time. In this limit, the restricting condition of "opaque" barrier of the nonrelativistic case is avoided, nevertheless, the very thin width of a single barrier to obtain this "transparent" limit can result in a problem itself, for probing the effect. A combination of two successive transparent barriers can show an apparent "superluminal" behavior  along a macroscopic arbitrary distance "L". Two solutions for energy $E$ above and below the potential square barrier $V$ are found, for both solutions there the apparent superluminal behavior is possible above a threshold of free travelling group velocity (energy) and dependent on the ratio  barriers length - free path as function of  the ratio group velocity - speed of light.
\end{abstract}

\pacs{03.65.Xp}

\maketitle

\section{INTRODUCTION}
The traversal time of a particle or a wave packet through a forbidden potential barrier \cite{har62,but82,but83} has not a unique definition both in nonrelativistic \cite{hau89} and relativistic case \cite{win04}. Different definitions of traversal times have been introduced, the most straightforward being the so called phase time.  The phase time  is defined, in the stationary-phase approximation, as  the energy derivative of the transmission phase shift: $\tau_p=\hbar d\alpha/dE$ given the transmission coefficient as $T(k)=\sqrt{T}e^{\imath\alpha}$ \cite{hau89}. Applying this definition, in the nonrelativistic case, to tunneling through a rectangular potential barrier of height $V_0>E$ and width "a" the phase time for a wave packet tends, in the limit of "opaque" barrier ($ka\gg 1$) to a constant value independent from the width "$a$" (Harman effect \cite{har62}) so that it could lead to apparent superluminal velocities. altough the interpretation of this apparent superluminal effect, is not a subject of the present article, the "opaque" barrier case is a very restricting condition considering the exponential decay of the amplitude through a tunneling process. In the case of relativistic particles \cite{chu02} through a barrier of width "$a$" the phase time has a different expression but it may be still recognized, in this case, a generalized Hartman effect. For example, in the case of two successive barriers \cite{lun07}, the phase time becomes independent, in the limit of "opaque" barrier, both from the width "a" and from the distance $L$ between the barriers. The limit of opaque barrier gives, by definition, strong constraints for  experimental probes, because the wave function amplitude decreases exponentially and this is more effective when more than one barrier is considered. For relativistic particles indeed, differently from the case of nonrelativistic particles, it is possible to consider the opposite limit of "transparent" barrier $ka\ll 1$ that leads, as well, to an apparent superluminal result for the phase time \cite{ger15}. In this article this limit of "transparent" barrier for relativistic particles is applied to a double barrier configuration. The presence of  two barriers of width $a$ and distance $L$, in some conditions, leads to a more evident apparent superluminal behavior where the ratio $a/L$ is a key factor.

\section{PHASE TIME IN THE APPROXIMATIONS OF "TRANSPARENT BARRIERS" AND RELATIVISTIC PARTICLES}
The  equations of the momentum outside ($\hbar k$) and inside ($\hbar q$) a potential barrier of height $V_0$ of a particle of mass $m$ and energy $E$, are
\begin{eqnarray}
\hbar kc=&& \sqrt{E^{2}-m^{2}c^{4}}\\
\hbar qc=&& \sqrt{m^{2}c^{4}-(V_0-E)^{2}}.
\end{eqnarray}
To have a proper tunneling, $V_0$ must be in the range $E-mc^{2}<V_0<E+mc^{2}$ because, below the lower limit the particle has enough energy to propagate over the potential barrier while, above the upper limit, the barrier can become supercritical and spontaneously emit positrons and electrons in the so called Klein tunneling \cite{dom00}. In the limit of "transparent" barriers ($qa\ll 1$), the potential satisfies two solutions: for $V_0$ greater than the total energy $E$ we have solution (a)
\begin{equation}
\label{sol1}
V_0\approx E+mc^{2}-\frac{(\hbar q)^{2}}{2m}\;\;\;\; \textrm{for}\;\;\; V_0>E
\end{equation}
and for $V_0$ lesser than the total energy $E$ we have solution (b)
\begin{equation}
\label{sol2}
V_0\approx E-mc^{2}+\frac{(\hbar q)^{2}}{2m} \;\;\;\; \textrm{for}\;\;\; V_0<E.
\end{equation}
The expression for the phase time across two potential barriers of width $a$ separated by a vacuum path of length $L$ is, from Lunardi et. al. \cite{lun07}
\begin{equation}
\label{ptime}
\tau_{p}=\frac{1}{\hbar c^{2}}\left\{(kL)\frac{E}{k^{2}}-\frac{1}{k^{2}q^{2}}\frac{h_{1}}{\Gamma^{2}+\Delta^{2}}\right\}.
\end{equation}
The expressions for $h_1$, $\Gamma$ and $\Delta$ are given in appendix. The approximation for "transparent" barriers ($qa\ll1$), at first order, is given by
\begin{eqnarray}
\label{rap}
&&\frac{h_{1}}{\Gamma^{2}+\Delta^{2}}\approx \left[(V_0-E)k^{2}\left(\frac{1}{\alpha}-\alpha\right)\right]+\\
&&\left[ -mc^{2}(k^{2}+q^{2})\left(\frac{1}{\alpha}+\alpha\right)\right]qa+O[qa]^{2}\nonumber
\end{eqnarray}
where $\alpha\equiv\frac{k}{q}\frac{(E-V_0+mc^{2})}{(E+mc^{2})}$. In the approximation of relativistic particles ($E\gg mc^{2}$) and "transparent" barriers ($qa\ll1$) solutions (a) (\ref{sol1}) and (b) (\ref{sol2}) are given in the following:

\subsection{Solution (a) for $E<V_0<E+mc^{2}$ }
For this solution with $V_0$ greater than $E$, $\alpha\approx \frac{\hbar kq}{2mE}\ll 1$. Substituting $V_0$ (\ref{sol1} into (\ref{rap})
\begin{equation}
\frac{h_{1}}{\Gamma^{2}+\Delta^{2}}\approx\left[\left(-\frac{\hbar^{2}q^{2}k^{2}}{2m}-mc^{2}q^{2}\right)\frac{1}{\alpha}\right]qa
\end{equation}
that, substituting $\alpha\approx \frac{\hbar kq}{2mE}\ll 1$  becomes
\begin{equation}
\frac{h_{1}}{\Gamma^{2}+\Delta^{2}}\approx\left(-Ekq-\frac{2m^{2}c^{2}E}{\hbar^{2}}\frac{q}{k}\right)qa;
\end{equation}
so, the phase time $\tau_{p}$ (\ref{ptime}) for this solution becomes
\begin{equation}
\tau_{p}\approx \left(\frac{L}{c^{2}}+\frac{a}{c^{2}}+\frac{2m^{2}}{\hbar^{2}k^{2}}a\right)\frac{E}{\hbar k}.
\end{equation}
Since the usual phase velocity of the free particle is $V_{\phi}=E/(\hbar k)$ the final expression for the phase time for this solution is
\begin{equation}
\label{tau1}
\tau_{p}\approx \frac{V_{\phi}}{c^{2}}\left[ L+a\left( 1+\frac{2m^{2}c^{2}}{\hbar^{2}k^{2}}\right)\right].
\end{equation}
\subsection{Solution (b) for $E-mc^{2}<V_0<E$}
For this solution with $E$ greater than $V_0$, $\alpha\approx \frac{2mc^{2}}{E} \frac{k}{q}\gg 1$. substituting $V_0$ (\ref{sol2} and $\alpha$ into (\ref{rap})
\begin{equation}
\frac{h_{1}}{\Gamma^{2}+\Delta^{2}}\approx \left(-\frac{c^{2}q \hbar^{2}k^{3}}{E}-\frac{2m^{2}c^{4}qk}{E}\right)qa,
\end{equation}
finally the expression of the phase time for this solution is
\begin{equation}
\label{tau2}
\tau_{p}\approx \frac{V_{\phi}}{c^{2}}\left[ L+\frac{c^{2}}{V_{\phi}^{2}} a\left( 1+\frac{2m^{2}c^{2}}{\hbar^{2}k^{2}}\right)\right]
\end{equation}
that is very similar to (\ref{tau1}) considering that, for relativistic particles, $V_{\phi}\simeq c$.

\section{Conditions for traversal time superluminal behavior}
For a free relativistic particle, $c^{2}/V_{\phi}=V_g$ where $V_g$ is the group velocity or the so called classical velocity of the particle; then,
 $\tau_f$ could be assumed as the  time it would take  a free relativistic particle to travel the same path of the example, i.e.
 \begin{equation}
 \tau_f=\frac{V_{\phi}}{c^{2}}\left(L+2a\right).
 \end{equation}

 \subsection{Conditions for solution (a)}
 So a free particle takes a longer time to travel the distance $L+2a$, than the phase time,  by an amount $\Delta t$
 \begin{equation}
 \label{tdiff}
 \Delta t\equiv\tau_f-\tau_{p}=\frac{V_{\phi}}{c^{2}}a\left[1-\frac{2m^{2}c^{2}}{\hbar^{2}k^{2}}\right]=\frac{a}{V_g}\left[3-2\frac{c^2}{V_g^2}\right]
 \end{equation}

because $c^2/V_g^2=(\hbar^2k^2+m^2c^2)/\hbar^2k^2$. The tunneling thus is a kind of accelerator of the motion. It must be recalled that equation (\ref{tdiff}) is valid for a relativistic particle with $V_g\simeq c$ and it can be seen from (\ref{tdiff}) that the time gain  of a tunneling relativistic particle with respect to a free particle begins when the velocity is $V_g>\sqrt{\frac{2}{3}}\: c=0.82 \: c $ and reaches the limit of  $a/V_g$ as $V_g$  grows toward the limiting value $c$.\\
The time gain could be such that the motion could be defined superluminal in the sense considered by the Hartman effect: defining the traversal velocity $V_T$ as the traveled path $L+2a$ divided by the phase time $\tau_p$, then
 \begin{equation}
 V_T=\frac{L+2a}{\tau_p}=\frac{L+2a}{\tau_f-\Delta t}\simeq \frac{L+2a}{\tau_f}+\frac{L+2a}{\tau_f^{2}}\Delta t
 \end{equation}
 that, in terms of free propagating group velocity $V_g$ and  of barriers length $a$, becomes
 \begin{equation}
 \label{traversal}
 V_T=V_g+V_g\frac{a}{L+2a}\left[ 3-2\frac{c^{2}}{V_{g}^{2}}\right].
 \end{equation}
 Let's now consider the conditions on $V_g$ and $a$ such that the traversal velocity $V_T$ tends toward the speed of light in vacuum $c$. Setting  $V_T\rightarrow c$, $V_g=\beta c$ and $a=\alpha L$, the (\ref{traversal}) becomes
\begin{equation}
\label{fig1}
\alpha=\frac{\beta^{2}-\beta}{2+2\beta-5\beta^{2}}.
\end{equation}

In figure (\ref{fig1}) $\alpha$ vs $\beta$ is plotted. The curve ($a$) shows the values for which $V_T\rightarrow c$. The region on the right of the curve is the region of superluminality. There is no solution for $\beta=(1+\sqrt{11})/5=0.8633$ so there is no superluminal effect for $V_g\leq 0.8633 c$, whatever be the barrier length $a$.

\begin{figure}
\includegraphics[width=8.6cm]{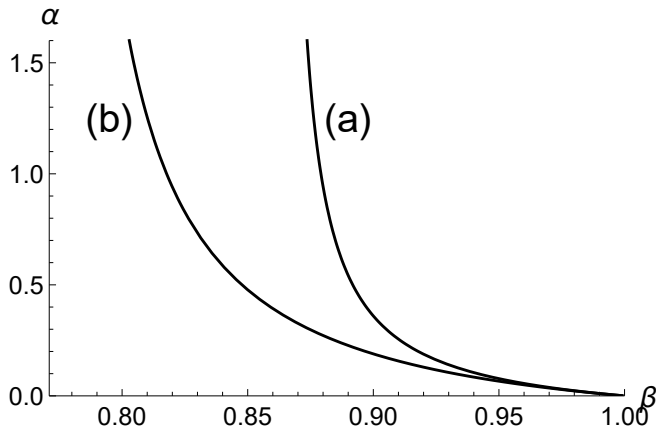}\\
  \caption{The curves show values of $\alpha\equiv a/L$ and $\beta\equiv V_g/c$ for which the traversal velocity $V_T\rightarrow c$. the region on the right of the curves is superluminal while on the left is subluminal. The curve (a) is for the energy $E$ lower than the potential barrier $V_0$ (case a), while the curve (b) is for the case where the energy $E$ is greater than the potential barrier $V_0$ (case b).}
  \label{figure1}
\end{figure}
Conversely, for $V_g\geq 0.8633 c$ there are values of $\alpha\equiv a/L$ for which $V_T\geq c$.

 \subsection{Conditions for solution (b)}
In the case in which $E>V_0>E-mc^2$, the gain in time is
\begin{equation}
\triangle t\equiv t_f-t_p=\frac{aV_g}{c^2}
\end{equation}
so the traversal velocity becomes
\begin{equation}
V_T=V_g+\frac{a}{L+2a}\frac{V_g^3}{c^2},
\end{equation}
and, differently from the previous case, the traversal velocity $V_T$ is always greater than the group velocity $V_g$.
Proceeding then like in the case (a), defining the ratios $\alpha=a/L$ and $\beta=V_g/c$ and setting the limit $V_T\rightarrow c$, the corresponding of equation (\ref{fig1}) is, in this case,
\begin{equation}
\alpha=\frac{1-\beta}{\beta^3+2\beta-2}.
\end{equation}
There is not a possible apparent superluminal behavior of the traversal velocity for $\alpha\leq 0$ thus for $\beta\leq [(9+\sqrt{105})^{2/3}-2\cdot 3^{1/3}]/[3^{2/3}(9+\sqrt{105})^{1/3}]$ i.e. for $ V_g\leq 0.7709 c$ while for values above this limit, the apparent superluminal behavior is represented by the space on the right of the curve ($b$) in Fig. \ref{figure1}.
\section{Conclusions}
For relativistic particles passing through a two forbidden barriers of width $a$ and distance $L$ in regime of "transparent" barrier approximation, $ka\ll 1$ an apparent superluminal behavior is found defining the traversal velocity as path divided by phase time. In the two cases of energy slightly above and under the barrier potential height, thresholds and conditions for superluminal behavior are found with the former case more favorable than the latter, with equivalent conditions, depending, the gain in time,  on the energy of the particle and proportional to the width $a$ of the barriers.

\appendix*
\section{}
\begin{equation}
\Gamma\equiv 8\alpha^{2}\cosh(2qa)-4(1+\alpha^{2})^{2}\sin^{2}(kL)\sinh^{2}(qa).
\end{equation}
\begin{equation}
\Delta\equiv 4\alpha (1-\alpha^{2})\sinh (2qa)+2(1+\alpha^{2})^{2}\sin(2kL)\sinh^{2}(qa).
\end{equation}
\begin{eqnarray}
&&h_1\equiv \Delta\{2(1+\alpha^{2})[(1+\alpha^{2})Eq^2(2kL)\sin(2kL)+\\
&&-4\alpha^2 mc^2(k^2+q^2)\cos(2kL)]\sinh^{2}(qa)-4\alpha^2 mc^2(k^2+q^2)\nonumber\\
&&[(1+\alpha^2)+(3-\alpha^2)\cosh(2qa)]+k^2(2qa)E-V_0)\nonumber\\
&&[(1+\alpha^2)^2\cos(2kl)-(1-6\alpha^2+\alpha^4)]\sinh(2qa))\}+\nonumber\\
&&+\Gamma \{-4\alpha(1-\alpha^2)k^2(2qa)(E-V_0)\cosh(2qa)+2(1+\alpha^2)\nonumber\\
&&[(1+\alpha^2)Eq^2(2kL)\cos(2kL)+4\alpha^2mc^2(k^2+q^2)\sin(2kL)]\nonumber\\
&&\sinh^2(qa)+[4\alpha(1-3\alpha^2)mc^2(k^2+q^2)-(1+\alpha^2)^2\nonumber\\
&&  k^2(2qa)(E-V_0)\sin(2kL)]\sinh(2qa)\}\nonumber
\end{eqnarray}

\end{document}